\documentclass[12pt]{article}
\usepackage[xdvi]{graphicx}
\makeatletter
\@addtoreset{equation}{section}
\makeatother

\begin{document}
\begin{titlepage}
\begin{flushright}
HIP-2007-02/TH \\
hep-th/0701159
\end{flushright}

\vspace*{2cm}

\begin{center}
{\large\bf 
Dependence of SU(N) coupling behavior on  
the size of extra dimensions
}
\end{center}

\vspace{1.5cm}

\begin{center}
{\bf Nobuhiro Uekusa}
\footnote{E-mail: nobuhiro.uekusa@helsinki.fi}
\end{center}

\vspace{0.2cm}

\begin{center}
{\it High Energy Physics Division, 
Department of Physical Sciences, University of Helsinki  
and Helsinki Institute of Physics, 
P.O. Box 64, FIN-00014 Helsinki, Finland}
\end{center}

\vspace{1cm}

\begin{abstract}
Coupling constants 
at high energy scales
are studied in SU(N) gauge theory with distinct sizes of extra dimensions.
We present the solution of gauge couplings as functions of the energy 
in such a way as to track the number of Kaluza-Klein modes.
In a flat extra dimension, it is shown that the gauge couplings 
have logarithmic dependence on the size of the extra dimension
and linear dependence on the number of Kaluza-Klein modes.
We find some patterns 
of the dependence on flavor of bulk and brane
fermions. 
Dependence of gauge couplings on the size of an extra dimension
is discussed also in a warped extra dimension.

\end{abstract}
\end{titlepage}
\newpage
\section{Introduction}

There have been many attempts to study 
the effect of extra dimensions.
Initial proposals for large extra dimensions
\cite{Antoniadis:1990ew} and 
warped extra dimensions \cite{Randall:1999ee}
assume that all particles in the standard model are confined on 
a brane with four-dimensional world volume.
For extra dimensions with the size smaller than the TeV scale,
it is an interesting possibility
that some fields are 
propagating in higher-dimensional bulk spacetime.
In such various brane-world scenarios,
it is possible for extra dimensions to have an effect
far below unification scale or the Planck scale.

One of the remarkable extra-dimensional effects
in brane-world scenario with bulk fields
is in running of coupling constants.
Higher-dimensional theory is non-renormalizable.
However, it can be assumed that the contributions from the Kaluza-Klein (KK) 
states with masses larger than the scale of interest are 
decoupled from the theory, 
giving rise to an approximately renormalizable theory.
In this approximation the corrections to the gauge couplings
can be calculated.
In case of flat extra dimensions,
it has been shown that the appearance of extra dimensions accelerates
the running of the gauge couplings,
due to the power-law corrections \cite{Dienes:1998vh, Dienes:1998vg}.
This has been studied in various models such as grand unification
\cite{Kobakhidze:2001yk, Hall:2001xb, Hall:2002ci,
Chaichian:2002uy, Kobakhidze:2006zq}.
Possible solutions to 
the problem raised on non-calculable corrections from 
unknown ultraviolet physics
have also been proposed in some scenarios 
\cite{Nomura:2001tn, Hebecker:2002vm}.

More recently, the running of gauge couplings has been studied 
in the universal extra dimension scenario where the extra dimension
is accessed by all the standard model fields 
\cite{Appelquist:2000nn, Macesanu:2002db}.
In \cite{Bhattacharyya:2006ym},
it has been shown that the size of the extra dimension is 
constrained in order that
the gauge couplings remain perturbative up to the scale where
they tend to unify.
This reveals the size of extra dimension to be regarded as 
a significant ingredient for running of gauge couplings
in brane-world scenario.

In this paper, 
we will examine how coupling constants in SU(N) gauge theory
behave at high energy scales, 
depending on the sizes of extra dimensions,
where we will take into account not only bulk fields
but also brane fields.
Our approach is based on four-dimensional KK effective
Lagrangian similar to \cite{Bhattacharyya:2006ym}.
The first KK excitation occurs at
the scale $(\pi/l)$ where $l$ is the size of the extra dimension.
Up to this scale the gauge coupling evolution 
has a contribution only from zero mode.
Between $(\pi/l)$ and $(2\pi/l)$, 
the running is still logarithmic but 
beta functions are modified due to the first KK excitation.
Whenever a KK threshold is crossed, 
beta functions are renewed.
In such a way as to track the number of KK states,
we will solve gauge couplings as functions of the energy 
using the background field method.
While the problem of ultraviolet sensitivity 
is beyond the scope of this paper,
our formulation will include possible localized kinetic terms.

We will also examine dependence of 
gauge couplings on the size of extra dimension
in Randall-Sundrum spacetime.
One-loop beta function of 
gauge couplings can be calculated in
a five-dimensional regularization scheme with holographic guide
\cite{Randall:2001gb}.
Equations in such a calculation
involve the curvature of $AdS_5$ denoted by $k$.
As suggested in certain models \cite{Maru:2006id,Maru:2006ji},
the size of extra dimension can be stabilized in the unit of $k$.
From this point of view, 
we will interpret $k$-dependent gauge couplings to
be $l$-dependent and
find an asymptotic 
solution for a gauge coupling in a form 
dependent on the size of extra dimension.

The paper is organized as follows.
In Sec. \ref{5dflat}, we will give Lagrangian for 
gauge fields and fermions in five-dimensional flat space
and derive four-dimensional KK effective Lagrangian.
After two kinds of gauge invariance in the theory will be shown,
we will solve gauge couplings using the background field method.
In Sec. \ref{5dwarped}, we will 
give an asymptotic 
solution for a gauge coupling
dependent on the size of extra dimension.
Summary and discussions will be given
in Sec. \ref{summary}.

\section{A flat extra dimension \label{5dflat}}

In this section, we will examine gauge couplings
on the orbifold $S^1/Z_2$.  
Our notation assumes
that gauge group SU(N) in four dimensions leaves unbroken,
that all fermions are Dirac fermions
and that there are  $2N_f$ bulk fermions and
$n_f$ brane fermions.

In this paper, capitalized indices $M,N$ run over 0,1,2,3,5,
lower-case indices $m$ run over 0,1,2,3 and
$(0),(n)$ are zero mode and KK mode indices where $n$ is 
a natural number.
We use a timelike metric $\eta_{mn}=\textrm{diag}(1,-1,-1,-1)$
and take the following basis for the Dirac matrices
\begin{eqnarray}
  \gamma^M=\left(\left(\begin{array}{cc}
		 0&\sigma^m \\
                 \bar{\sigma}^m& 0\\
		       \end{array}\right),
          \left(\begin{array}{cc}
	   -i&0 \\
            0&i \\
		\end{array}\right)\right) ,
\end{eqnarray}
where $\sigma^m=(1,\overrightarrow{\sigma})$,
$\bar{\sigma}^m=(1,-\overrightarrow{\sigma})$.

\subsection{Lagrangian and gauge invariance}

We consider five-dimensional SU(N) gauge theory 
on the orbifold $S^1/Z_2$ with the metric
\begin{eqnarray}
  ds^2=g_{MN}dx^M dx^N= \eta_{mn} dx^m dx^n - dx^5 dx^5 .
\end{eqnarray}
Two branes are located at $x^5=0,l$.
The five-dimensional starting Lagrangian is  
\begin{eqnarray}
  {\cal L}= {\cal L}_{\textrm{\scriptsize gauge}} 
    +{\cal L}_{\textrm{\scriptsize fermion}} .
\end{eqnarray}
The Lagrangian 
composed purely of the gauge field 
$A_M(x,x^5)=A_M^a(x,x^5) t^a$ is written as
\begin{eqnarray}
 {\cal L}_{\textrm{\scriptsize gauge}}=-{1\over 4g^2} (F_{MN}^a)^2
  -{1\over 4g_1^2} (F_{mn}^a {}_{(0)})^2\delta(x^5)
  -{1\over 4g_2^2} (F_{mn}^a {}_{(0)})^2\delta(x^5-l) , \label{lgauge}
\end{eqnarray}
with the field strength 
 $F_{MN}^a=\partial_M A_N^a-\partial_N A_M^a
    +f^{abc}A_M^b A_N^c $,
where  $g,g_1,g_2$ are the gauge coupling constants
and $f^{abc}$ is the structure constant.
The localized kinetic terms are denoted as 
the terms multiplied by the delta function.
The Lagrangian for one-flavor bulk fermions is given by
\begin{eqnarray}
 {\cal L}_{\textrm{\scriptsize fermion}} 
=\bar{\psi}^i (i\gamma^M D_M)\psi^i ,
\end{eqnarray}
with $i=1,2$.
The covariant derivative is
 $  D_M=\partial_M-iA_M^a t^a $.

Let us derive the four-dimensional Lagrangian 
so that gauge coupling of zero mode can be calculated.
By KK mode expansion, the gauge field is decomposed  into
\begin{eqnarray}
   &&A_m^a(x,x^5)=
       A_m^a{}_{(0)}(x) 
  +\sum_{n=1} \sqrt{2} A_m^a{}_{(n)}(x) 
   \cos({n\pi x^5\over l})
\\
   &&A_5^a(x,x^5)=\sum_{n=1}
   \sqrt{2} A_5^a{}_{(n)}(x) 
  \sin({n\pi x^5\over l}) .
\end{eqnarray}
Substituting the KK mode expansion into the Lagrangian (\ref{lgauge}),
we obtain the four-dimensional gauge part Lagrangian
\begin{eqnarray}
 \int_0^l ~dx^5~ {\cal L}_{\textrm{\scriptsize gauge}}
 &\!\!=\!\!&-{1\over 4g^2} 
  (l+{g^2\over g_1^2}+{g^2\over g_2^2})
   (F_{mn}^a {}_{(0)})^2
\nonumber
\\
&&
 -{l\over 4g^2} 
 \sum_{n=1}\left((F_{mn}^a {}_{(n)})^2
  +2 f^{abc}F_{mn}^a{}_{(0)}  A^{mb}_{(n)} A^{nc}_{(n)}
  +2F_{m5}^a {}_{(n)}F^{m5a} _{(n)}\right) .
\nonumber
\\
&&
  \label{lag4d} 
\end{eqnarray}
The field strengths are given by
\begin{eqnarray}
  F_{mn}^a{}_{(0)}&\!\!=\!\!&
 \partial_m A_n^a{}_{(0)}- \partial_n A_m^a{}_{(0)}
 +f^{abc}A_m^b{}_{(0)}A_n^c{}_{(0)} ,
\nonumber
\\
  F_{mn}^a{}_{(n)}&\!\!=\!\!&
 \partial_m A_n^a{}_{(n)}- \partial_n A_m^a{}_{(n)}
 +f^{abc}A_m^b{}_{(m)}A_n^c{}_{(r)}\delta_{nmr}^+
 +f^{abc}(A_m^b{}_{(0)}A_n^c{}_{(n)}-A_n^b{}_{(0)}A_m^c{}_{(n)}) ,
\nonumber
\\
  F_{m5}^a {}_{(n)} &\!\!=\!\!&
  \partial_m A_5^a {}_{(n)}
   +({n\pi\over l})A_m^a {}_{(n)}
   +f^{abc} A_m^b {}_{(m)} A_5^c{}_{(r)}
    \delta_{nmr}^-
   +f^{abc}A_m^b{}_{(0)}A_5^c{}_{(n)} , 
\nonumber
\\
  &&\label{fm5}
\end{eqnarray}
with
 $\delta_{nmr}^{\pm}
 =(\delta_{n,m+r}+\delta_{m,n+r}\pm\delta_{r,n+m})/\sqrt{2}$.
From Eqs.(\ref{lag4d}) and (\ref{fm5}), it is seen that
the $n$-th KK gauge field
$A_m^a{}_{(n)}$ has the mass $(n\pi/l)$.
The $n$-th KK gauge field begins to play a dynamical role 
at energy scales higher than $(n\pi/l)$. 
We treat the summation over KK modes in the Lagrangian (\ref{lag4d})
as scale-dependent.
This is explicitly written as $\sum_{n=1}^m$ 
at the energy range $m\pi/l \le E< (m+1)\pi/l$.
At scales less than $(\pi/l)$,
the Lagrangian describes  only zero mode
as the summation is simply zero.
The $n$-th KK scalar $A_5^a{}_{(n)}$ has the same mass as that of 
the $n$-th KK gauge field, 
as more explicitly seen after gauge fixing.
The summation over the KK modes is treated
similarly to the KK gauge field case.
 
The four-dimensional Lagrangian (\ref{lag4d})
is invariant under a standard four-dimensional 
gauge transformation.
The infinitesimal gauge transformation law for $A_m^a {}_{(0)}$
is given by  
\begin{eqnarray}
  \delta A_m^a {}_{(0)}
  =\partial_m \alpha^a 
    +f^{abc} A_m^b {}_{(0)} \alpha^c  
\end{eqnarray}
where $\alpha^a(x)$ is an infinitesimal parameter
dependent on four-dimensional coordinates.
The KK gauge field and scalar are transformed as 
adjoint matter fields,
\begin{eqnarray}
 \delta A_m^a {}_{(n)}&\!\!=\!\!&f^{abc}A_m^b{}_{(n)} \alpha^c ,
 \label{eq:trsfm}
\\
 \delta A_5^a {}_{(n)}&\!\!=\!\!&f^{abc}A_5^b {}_{(n)}\alpha^c .
 \label{eq:trsf5}
\end{eqnarray}
In addition to the standard gauge invariance,
the Lagrangian (\ref{lag4d})
is also invariant under another gauge transformation
with the infinitesimal local parameter $\alpha^a_{(n)} (x)$, 
\begin{eqnarray}
  &&\delta_{(n)} A_m^a {}_{(0)}=0 ,
\\
  &&\delta_{(n)} A_m^a {}_{(n)}
  =\partial_m \alpha^a_{(n)} 
    +f^{abc} A_m^b {}_{(m)} \alpha^c_{(r)} \delta_{nmr}^+  
    +f^{abd} A_m^b {}_{(0)} \alpha^c_{(n)} ,
\\
 &&\delta_{(n)} A_5^a {}_{(n)}
  =-({n\pi\over l})\alpha_{(n)}^a
   +f^{abc}A_5^b {}_{(m)}\alpha^c_{(r)}\delta_{nmr}^- .
\end{eqnarray}
Using these two kinds of gauge invariance,
we will choose gauge fixing convenient for the background field method.  

Let us consider fermionic part.
In order to calculate running of gauge coupling constant,
we concentrate on quadratic fermionic Lagrangian. 
The $Z_2$ projection for the fermion fields is defined as
\begin{eqnarray}
 \left\{
  \begin{array}{l}
   \psi^1(x,-x^5)=i\gamma_5\psi^1(x,x^5) ,\\
     \psi^1(x,-x^5+l)=i\gamma_5\psi^1(x,x^5+l) , \\
  \end{array}
 \right. 
 ~~
 \left\{
  \begin{array}{l}
   \psi^2(x,-x^5)=-i\gamma_5\psi^2(x,x^5) ,\\
     \psi^2(x,-x^5+l)=-i\gamma_5\psi^2(x,x^5+l) . \\
  \end{array}
 \right.
\end{eqnarray}
From this parity assignment and the chiral representation
$\psi^i=(\phi^i_L,\phi^i_R)^T$, 
the fermion fields are decomposed into
\begin{eqnarray}
 \phi_L^1(x,x^5)&\!\!=\!\!& {1\over \sqrt{l}}\psi_L(x)
    +\sum_{n=1}\sqrt{2\over l}\phi_L^1{}_{(n)}(x)
  \cos({n\pi x^5\over l}) ,
\\
 \phi_R^1 (x,x^5)&\!\!=\!\!& 
  \sum_{n=1}\sqrt{2\over l}\phi_R^1{}_{(n)}(x)
  \sin({n\pi x^5\over l}) ,
\\
 \phi_L^2 (x,x^5)&\!\!=\!\!& 
 \sum_{n=1}\sqrt{2\over l}\phi_L^2{}_{(n)}(x)
  \sin({n\pi x^5\over l}) ,
\\
 \phi_R^2(x,x^5)&\!\!=\!\!& {1\over \sqrt{l}}\psi_R(x)
    +\sum_{n=1}\sqrt{2\over l}\phi_R^2{}_{(n)}(x)
  \cos({n\pi x^5\over l}) .
\end{eqnarray}
Writing four-dimensional Dirac fermions as
\begin{eqnarray}
 \psi=\left(\begin{array}{l}
       \psi_L\\ \psi_R
	    \end{array}\right) ,
  ~~
 \psi^i_{(n)}=\left(\begin{array}{l}
       \phi_L^i{}_{(n)}\\ \phi_R^i{}_{(n)} 
	    \end{array}\right) ,
\end{eqnarray}
we obtain four-dimensional fermionic Lagrangian 
\begin{eqnarray}
\int_0^l ~dx^5~ {\cal L}_{\textrm{\scriptsize fermion}}
 \bigg|_{\textrm{\scriptsize quad}}
&\!\!=\!\!&\bar{\psi} (i\gamma^m \partial_m)\psi
\nonumber
\\
&& +\sum_{n=1}\bar{\psi}^1_{(n)}
\left(i\gamma^m \partial_m-{n\pi\over l}\right)\psi^1_{(n)}
+\sum_{n=1}\bar{\psi}^2_{(n)}
 \left(i\gamma^m \partial_m+{n\pi\over l}\right)\psi^2_{(n)} .
\nonumber
\\
 \label{lag4df}
\end{eqnarray}
The summation over KK modes is treated as scale-dependent
similarly to the KK gauge and scalar.
The Lagrangian for $2N_f$ bulk fermions are obtained as 
the number of the Lagrangian (\ref{lag4df}) is $N_f$.
The Lagrangian for $n_f$ brane fermions are obtained as
the number of the first term 
in the right hand side of (\ref{lag4df}) is $n_f$.

\subsection{Solution of gauge couplings}

We have set up the four-dimensional effective Lagrangian with
a scale-dependent KK mode summation.
Now we apply the background 
field method to the Lagrangians 
(\ref{lag4d}) and (\ref{lag4df}).
The zero mode gauge field is decomposed into
classical background and quantum fluctuation 
\begin{eqnarray}
  A_m^a{}_{(0)} \to A_m^a{}_{(0)} +{\cal A}_m^a  .
\end{eqnarray}
The KK  gauge field and scalar are interpreted as
quantum fluctuation around zero classical background. 
Then the field strengths in (\ref{fm5}) 
become 
\begin{eqnarray}
  F_{mn}^a{}_{(0)}  &\!\!\to\!\!& 
  F_{mn}^a {}_{(0)}+
   D_m {\cal A}_n^a-D_n {\cal A}_m^a
    +f^{abc}{\cal A}_m^b {\cal A}_n^c ,
\nonumber
\\
  F_{mn}^a{}_{(n)} &\!\!\to\!\!& 
   F_{mn}^a{}_{(n)}+f^{abc}(A_m^b{}_{(n)}{\cal A}_n^c 
                        -A_n^b{}_{(n)}{\cal A}_m^c) , 
\\
  F_{m5}^a{}_{(n)} &\!\!\to\!\!& 
   F_{m5}^a{}_{(n)}-f^{abc}A_5^b{}_{(n)}{\cal A}_m^c ,
\nonumber
\end{eqnarray}
where $D_m$ is the covariant derivative with respect to 
the background gauge field $A_m^a{}_{(0)}$.
Since the four-dimensional Lagrangian has
gauge invariance for the transformation parameters
$\alpha^a$ and $\alpha_{(n)}^a$,
we need to introduce two types of gauge fixing in order to define
the functional integral.
For the parameter $\alpha^a$, the gauge fixing function can be chosen as
\begin{eqnarray}
 G^a=
 \sqrt{1+{g^2\over lg^2_1}+{g^2\over lg^2_2}}\,D^m {\cal A}_m^a 
  -\omega^a(x)
\end{eqnarray}
with a Gaussian weight for $\omega^a$.
The corresponding ghost and antighost are 
denoted as $c$ and $\bar{c}$, respectively.
For the other parameter $\alpha_{(n)}^a$, 
we choose the other gauge fixing function
so as to cancel $A_m{}_{(n)}$-$A_5{}_{(n)}$ mixing terms 
in kinetic Lagrangian,
\begin{eqnarray}
 G^a_{(n)}=
  D^m A_m^a{}_{(n)} -\left({n\pi\over l}\right)A_5^a{}_{(n)} 
  -\omega^a_{(n)},
\end{eqnarray}
with a Gaussian weight for $\omega^a_{(n)}$.
The corresponding ghost and antighost are
denoted as $c_{(n)}$ and $\bar{c}_{(n)}$, respectively.
With these two types of gauge fixing functions, 
total quadratic Lagrangian 
is obtained as
\begin{eqnarray}
 {\cal L}_{\textrm{\scriptsize quad}}
  ={\cal L}_{\cal A}+{\cal L}_n
   +{\cal L}_5+{\cal L}_c+{\cal L}_\psi ,
 \label{gflaa}
\end{eqnarray}
\begin{eqnarray}
 {\cal L}_{\cal A}
 &\!\!=\!\!& -{1\over 2g^2}
 \left(l+{g^2\over g_1^2}+{g^2\over g_2^2}\right)
   \left\{
    {1\over 2}(D_m {\cal A}_n^a-D_n {\cal A}_m^a)^2
 \right.
\nonumber
\\
 &&\qquad\qquad\qquad\qquad\qquad\qquad \left.
    +f^{abc}F^a_{mn}{}_{(0)}{\cal A}^{mb}{\cal A}^{nc}
  +(D^m {\cal A}_m^a)^2\right\} ,
\\
 {\cal L}_{n}
 &\!\!=\!\!&-{l\over 2g^2}\left\{
    {1\over 2}(D_m A_n^a{}_{(n)}-D_n A_m^a{}_{(n)})^2
    -\left({n\pi\over l}\right)^2 A_m^a{}_{(n)}A^{ma}_{(n)}
  \right. 
\nonumber
\\  &&\qquad\qquad\qquad\qquad\qquad\qquad
   \left.
   +f^{abc}F^a_{mn}{}_{(0)}A^{mb}_{(n)}A^{nc}_{(n)}
  +(D^m A_m^a{}_{(n)})^2\right\} ,
\\
 {\cal L}_{5}
 &\!\!=\!\!&{l\over 2g^2}\left\{
    (D^m A_5^a{}_{(n)})(D_m A_5^a{}_{(n)})
   -\left({n\pi\over l}\right)^2
   A_5^a{}_{(n)}A_5^a{}_{(n)}\right\}  ,
\\
 {\cal L}_{c}
&\!\!=\!\!&-i\sqrt{1+{g^2\over lg_1^2}+{g^2\over lg_2^2}}~
  \bar{c}^a (D^2)^{ac}c^c
  -i~
  \bar{c}^a{}_{(n)} (D^2+\left({n\pi\over l}\right)^2)^{ac}c_{(n)}^c ,
\\
 {\cal L}_{\psi}
&\!\!=\!\!&\bar{\psi} (i\gamma^m D_m)\psi
 +\bar{\psi}^1_{(n)}
 \left(i\gamma^m D_m-{n\pi\over l}\right)\psi^1_{(n)}
+\bar{\psi}^2_{(n)}
 \left(i\gamma^m D_m+{n\pi\over l}\right)\psi^2_{(n)} .
  \label{lpsi}
\end{eqnarray} 
with summation over KK modes and flavor.

\begin{figure}[htb]
\begin{center}
 \includegraphics[width=6.5cm]{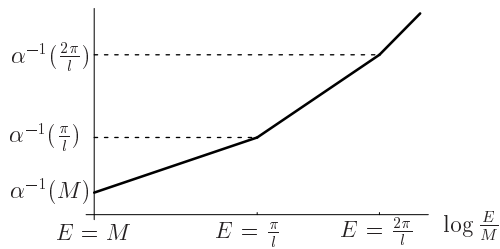} 
\caption{Behavior of a coupling is
 schematically drawn.
 \label{figa}}
\end{center}
\end{figure}

We now calculate  gauge coupling constants
in such a way to track the number of KK modes.
For scales less than the mass of the first KK mode,
the gauge coupling constant 
$\alpha^{-1}\equiv 4\pi g^{-2}(l+g^2 g_1^{-2}+g^2 g_2^{-2})$
is given by
\begin{eqnarray}
 \alpha^{-1}(E)= 
 \alpha^{-1}(M)+{b\over 2\pi} \log {M\over E} ,
 ~\quad \textrm{for}~~ E < {\pi\over l} ,
 \label{alpha0}
\end{eqnarray}
where $E$ and $M$ are the scale of interest and 
another scale. 
In Eq.(\ref{alpha0}),  $b$ is obtained only from zero mode part 
in the gauge-fixed Lagrangian (\ref{gflaa}) as
\begin{eqnarray}
  b= -\left({11\over 3}C_2(G)
     -{4\over 3}(N_f+n_f)C(r)\right) ,
\end{eqnarray}
with $C_2(G)=N$ and $C(r)={1\over 2}$.
The contributions to the factor $b$ from each field is tabulated
in Table~\ref{tabb}. 
As scales cross the mass of the first KK excitation, 
the first KK mode becomes dynamical.
Then $\alpha^{-1}$ is given by
\begin{eqnarray}
  \alpha^{-1}(E)=\alpha^{-1}(M)
   +{b\over 2\pi} \log {M\over E}
   +{b_{\textrm{\scriptsize KK}}\over 2\pi}
    \log {\pi\over El} ,
 ~\quad \textrm{for}~~
   {\pi \over l} \le E < {2\pi\over l} ,
\end{eqnarray}
where
\begin{eqnarray}
  b_{\textrm{\scriptsize KK}}
 =-\left({21\over 6}C_2(G)-{8N_f\over 3}C(r)\right) .
 \label{bkk}
\end{eqnarray}
The contributions to $b_{\textrm{\scriptsize KK}}$ from
each field is tabulated in Table~{\ref{tabb}}.
\begin{table}[htbp]
\begin{center}
\caption{Contributions to $b$ and $b_{\textrm{\scriptsize KK}}$
\label{tabb}}
\begin{tabular}{ccc} 
\multicolumn{3}{c}{(i) Zero mode contributes to $b$} \\
  \hline \hline
  Dirac   & gauge   & ghost  \\
  $\psi$ &  $A_m{}_{(0)}$ & $c,\bar{c}$ 
 \\ \hline 
 ${4(N_f+n_f)C(r)\over 3}$ & 
 $-{10C_2(G)\over 3}$ &
 $-{C_2(G)\over 3}$ \\ \hline \hline   
\end{tabular}
~~~\qquad
\begin{tabular}{cccc} 
 \multicolumn{4}{c}{(ii) KK mode contributes to 
$b_{\textrm{\scriptsize KK}}$} \\
  \hline \hline
   Dirac & gauge & scalar & ghost \\
   $\psi^i {}_{(n)}$ &
    $A_m{}_{(n)}$ & 
   $A_5{}_{(n)}$ &  $c_{(n)},\bar{c}{}_{(n)}$\\ \hline
 ${8N_f C(r)\over 3}$ &  
 $-{10C_2(G)\over 3}$ &
 ${C_2(G)\over 6}$ &
 $-{C_2(G)\over 3}$ \\ \hline \hline
\end{tabular} 
\end{center}
\end{table}
As scales are higher than the $n$-th KK mass,
the first $n$ KK modes become dynamical.
In this way, we find a general solution for $\alpha^{-1}$,
\begin{eqnarray}
  \alpha^{-1}(E)&\!\!=\!\!&\alpha^{-1}(M)
   +{b\over 2\pi} \log {M\over E}
   +{b_{\textrm{\scriptsize KK}}\over 2\pi}
    \sum_{m=1}^n\log {m \pi \over El} 
\nonumber
\\ 
 &\!\!=\!\!&\alpha^{-1}(M)
   +{b\over 2\pi} \log {M\over E}
   +{b_{\textrm{\scriptsize KK}}\over 2\pi}
    \left(n\log {\pi\over El}+\log n!\right)  ,
 ~ \textrm{for}~~
   {n\pi \over l} \le E < {(n+1)\pi\over l} .
\nonumber
\\
 \label{generalsol}
\end{eqnarray}
From this equation, it is seen that
the gauge coupling has a linear dependence on KK level $n$
and a logarithmic dependence on size of extra dimension $l$. 
Behavior of the $\alpha^{-1}$ is schematically shown in Fig~\ref{figa}.
Below the scale of the appearance of the extra dimension,
the gauge coupling has the contribution only from zero mode 
and the slope is $-b/(2\pi)$ for $\log E$.
At $E=\pi/l$ where the first KK excitation occurs,
the gauge coupling begins to have
the slope $-(b+b_{\textrm{\scriptsize KK}})/(2\pi)$.  
Whenever scales cross a KK mass, the new KK mode with the mass is added 
and at $n\pi/l \le E< (n+1)l/\pi$ the slope has 
$-(b+n b_{\textrm{\scriptsize KK}})/(2\pi)$.
Unless $|b|\gg |b_{\textrm{\scriptsize KK}}|$, 
the running of the gauge coupling is  strongly sensitive to 
the number of KK modes 
once the extra dimension appears with a size.
For example, if the size of extra dimension is $l^{-1}=10^5$ GeV$/\pi$,
the first 100 KK modes linearly would enhance the gauge coupling
at the scale of $10^7$ GeV$/\pi$.

\begin{figure}[htb]
\begin{minipage}{0.5\hsize}
  \begin{center}
   \includegraphics[width=8cm]{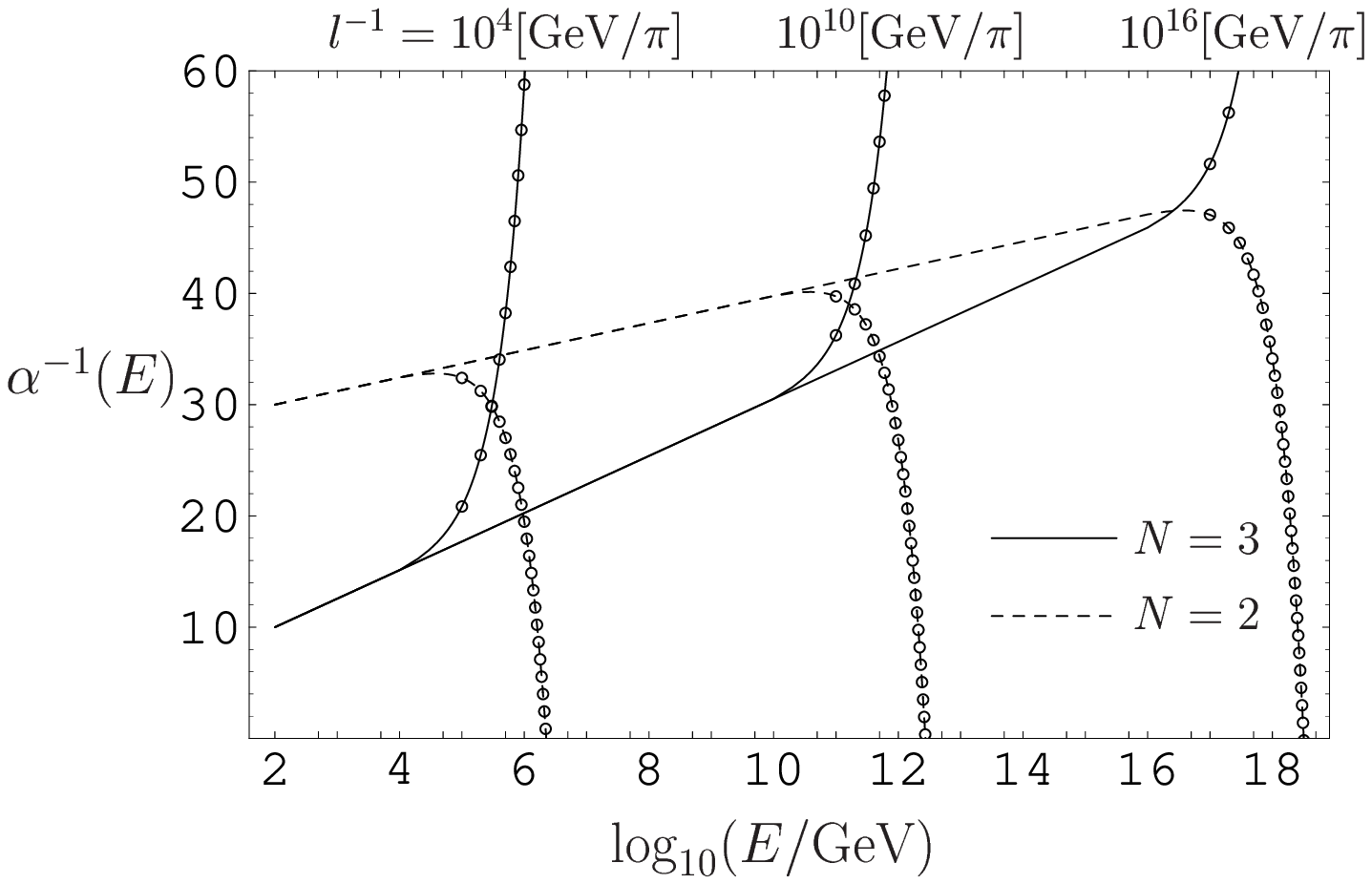} \\
(i)~ $(N_f, n_f)=(6,0)$.
  \end{center}
\end{minipage}
\begin{minipage}{0.5\hsize}
  \begin{center}
   \includegraphics[width=8cm]{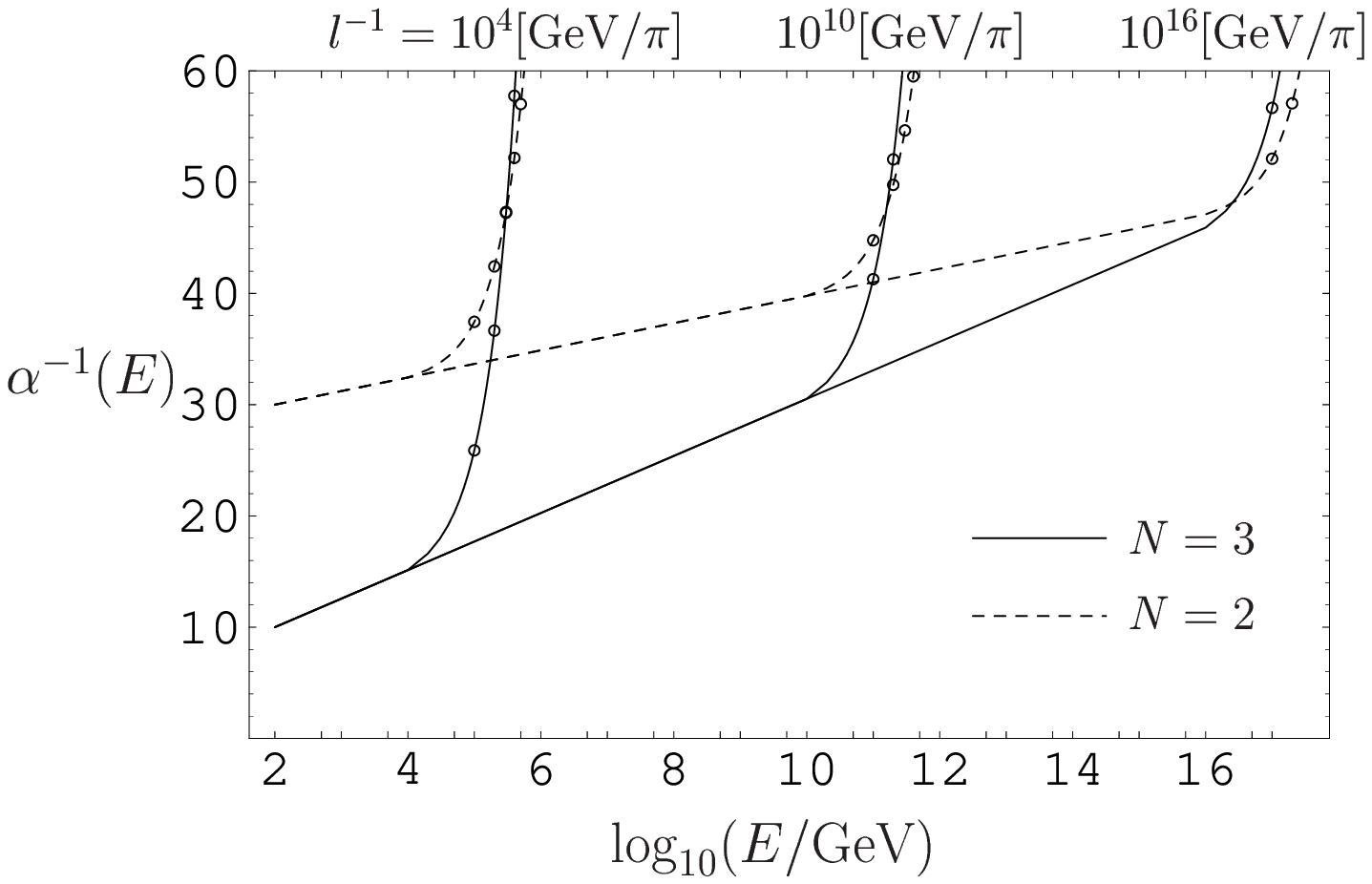} \\
  (ii)~ $(N_f, n_f)=(3,3)$.
  \end{center}
\end{minipage}

\vspace{7mm}

\begin{minipage}{0.5\hsize}
  \begin{center}
   \includegraphics[width=8cm]{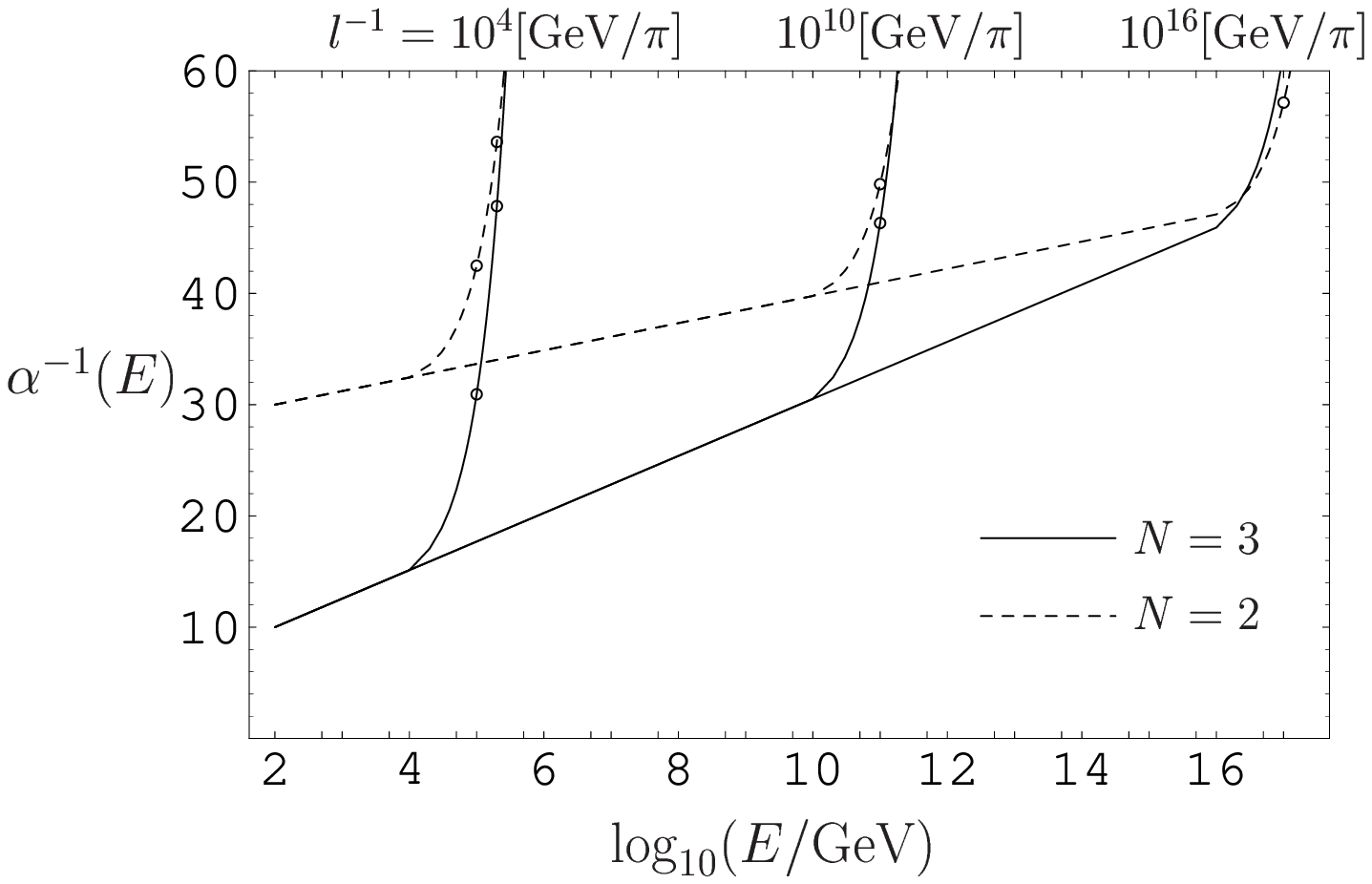} \\
    (iii)~ $(N_f,f_f)=(0,6)$.
  \end{center}
\end{minipage}
\caption{%
Behavior of the coupling as a function of the energy.
Circles show the appearance of the ($10n$)-th KK mode
where $n$ is a natural number.
At $E=M(=100\textrm{GeV})$, the coupling constants are chosen as
$\alpha^{-1}=10$ and 30 for $N=3$ and 2.
  \label{ael}}
\end{figure}

In order to find more specific patterns,
we choose the numbers of flavor and 
color as well as the size of extra dimension as
$N_f=6,3,0$, $n_f=6,3,0$, $N=2,3$ and
$l^{-1}=10^4, 10^{10}, 10^{16}$ [GeV$/\pi$].
And we fix the total number of flavor as $N_f+n_f=6$.
Behavior of the gauge coupling as
a function of the energy is shown in Fig.~\ref{ael}.
In the figure, the appearance of KK modes is shown with circles.
As mentioned above,
after the appearance of the extra dimension 
the gauge coupling increasingly evolves 
by including KK modes.
An indication of the coupling evolution can be
the energy when the gauge couplings for $N=2,3$ has the same value.
At this energy, there is the same number of KK modes 
for $N=2,3$ since 
KK masses are independent of color.
Defining the number of the KK modes as $n_*$
and the energy at a coincident point of gauge couplings as $E_*$,
we find from Eq.(\ref{generalsol})
\begin{eqnarray}
 E_*
 ={\pi\over l}
  \left((n!)^{\Delta b_k}
  \left({Ml\over \pi}\right)^{\Delta b}
  e^{2\pi \Delta \alpha^{-1}}
  \right)
   ^{1/\Delta b+n\Delta b_k} 
 ~ \textrm{for}~~
   {n_*\pi \over l} \le E_* < {(n_*+1)\pi\over l} ,
\end{eqnarray}
where $\Delta b_k=b_{\textrm{\scriptsize KK}\,N=2}
-b_{\textrm{\scriptsize KK}\,N=3}$,
$\Delta b=b_{N=2}-b_{N=3}$ and
$\Delta \alpha^{-1}=\alpha^{-1}(M)_{N=2}-\alpha^{-1}(M)_{N=3}$.
When the total number of flavor is fixed, 
dependence of $E_*$ on flavor occurs only due to $\Delta b_k$.
If $N_f$ with fixed ($N_f+n_f$)
is chosen as independent of color,
$\Delta b_k$ becomes independent of flavor.
Then the energy at a coincident point of gauge couplings 
is determined independently
the value of $N_f$.
In each case of $(N_f,n_f)$, $E_*$
is shown in Table~\ref{enj}.
\begin{table}[ht]
\begin{center}
\caption{Energies at the coincident points of the gauge couplings
 in the unit of GeV.
In each box, three quantities are denoted as the energies for
 $l^{-1}=10^4, 10^{10}, 10^{16}$[GeV/$\pi$] from above.
For three cases shown in left-bottom boxes there is no solution.
 \label{enj}}
 \includegraphics[width=9.5cm]{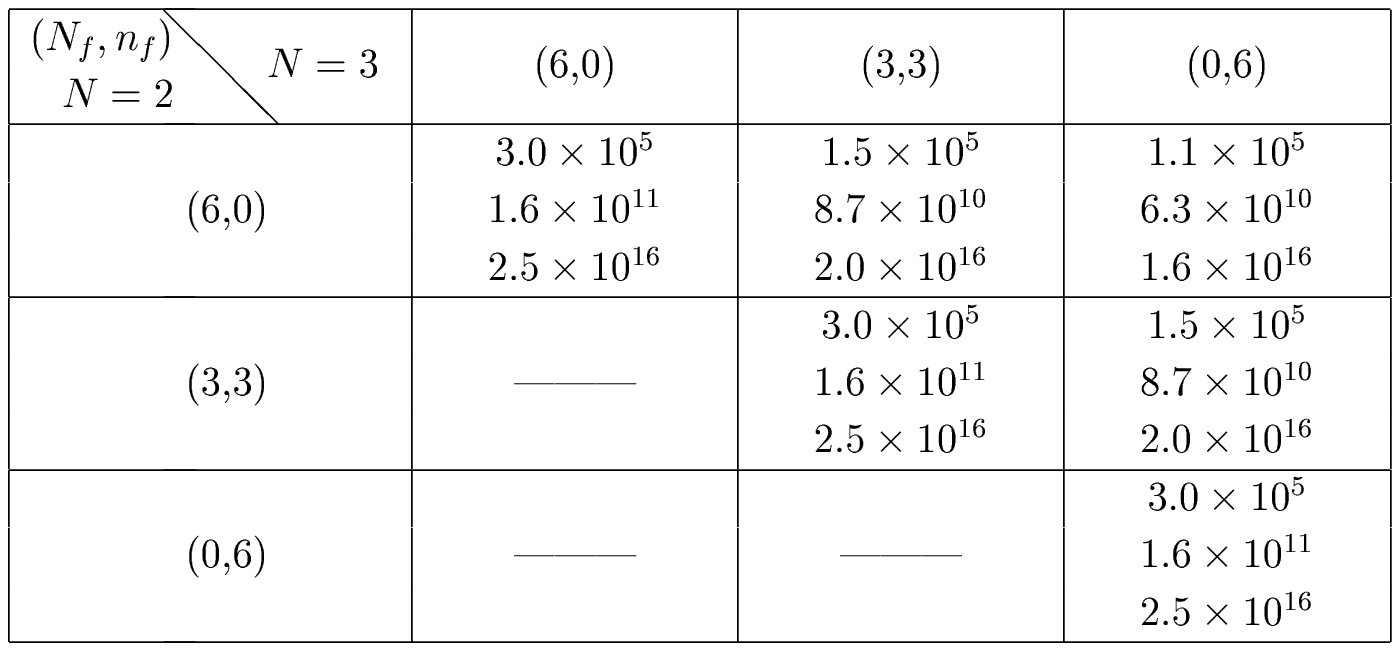} 
\end{center}
\end{table}
As seen also from Fig.~\ref{ael},
there is no solution in the case where
$N_f{}_{N=2} < N_f{}_{N=3}$. 
For $N_f{}_{N=2} \ge N_f{}_{N=3}$,
as $N_f{}_{N=3}$ increases,
the energy $E_*$ tends to increase.
This is because the relative sign 
between bosons and fermions in the $b_{\textrm{\scriptsize KK}}$ (\ref{bkk})
makes the evolution less for larger $N_f$.
For $N_f{}_{N=2} = N_f{}_{N=3}$,
$N_f$-value-independence of $E_*$ is
seen in diagonal boxes in Table~\ref{enj}.
In diagonal boxes
and upper right boxes along them,
it is also seen 
that $E_*$ is characterized by $(N_f{}_{N=2}-N_f{}_{N=3})$.
These features of $E_*$ can be read in the view of
the number of KK modes from
the fact that $n_*$ is a maximum integer less than $E_*l/\pi$.
The maximum number of $n_*$ is obtained for
$N_f{}_{N=2}=N_f{}_{N=3}$ and largest $N_f{}_{N=3}$.
From Table~\ref{enj},
it is seen that
the number of KK modes for coinciding of the couplings is at most
30, 16, 2 for $l^{-1}=10^4,10^{10},10^{16}$[GeV$/\pi$], respectively.
Similar behavior is obtained for 
more general two gauge groups.

\section{A warped extra dimension \label{5dwarped}}

In this section, we consider SU(N) gauge theory 
in the Randall-Sundrum geometry with metric
\begin{eqnarray}
 ds^2= g_{MN}dx^M dx^N= {1\over (kz)^2}(\eta_{mn} dx^m dx^n - dz^2) .
\end{eqnarray}
For simplicity, we consider purely gauge part
\begin{eqnarray}
 {\cal S}=\int d^4x dz \sqrt{-\textrm{det}(g_{MN})}
   \left(-{1\over 4g^2}(F_{MN}^a)^2\right) .
\end{eqnarray}
As mentioned in Introduction, 
one-loop beta function of 
gauge couplings can be calculated in
a five-dimensional regularization scheme.
With the background field method,
the gauge coupling is obtained as \cite{Randall:2001gb} 
\begin{eqnarray}
 \alpha^{-1}(E)=\alpha^{-1}(M)
     +{b_w\over 2\pi}\log {M\over E} ,
\end{eqnarray}
\begin{eqnarray}
   b_w=-C_2(G)\left({11\over 3}I^{1,0}(0,\Lambda/k)
     -{1\over 6}I^{1,i}(0,\Lambda/k)\right) ,
\end{eqnarray}
where 
\begin{eqnarray}
 I^{1,\nu}(y_1,y_2)
  =2\int_{y_1}^{y_2} y dy \int_y^{y_2} z dz
  \left(
  {(A K_\nu(y)+B I_\nu(y))(C K_\nu(z)+D I_\nu(z))
  \over AD-BC}\right)^2 ,
  \label{i0nu}
\end{eqnarray}
\begin{eqnarray}
  A&\!\!=\!\!& y_1 I_{\nu-1}(y_1)-(\nu-1)I_\nu(y_1) ,
\\
  B&\!\!=\!\!& y_1 K_{\nu-1}(y_1)-(\nu-1)K_\nu(y_1) ,
\\
  C&\!\!=\!\!& y_2 I_{\nu-1}(y_2)-(\nu-1)I_\nu(y_2) ,
\\
  D&\!\!=\!\!& y_2 K_{\nu-1}(y_2)-(\nu-1)K_\nu(y_2) .
\end{eqnarray}
Now we calculate asymptotic form of the gauge coupling for 
$k \ll \Lambda$.
The modified Bessel function $K_\nu$ is singular at $y_1\to 0$
so we will take limit $y_1\to 0$ after integral is performed.
Throughout this analysis, we take into account $y_1 \ll 1$ and $y_2\gg 1$.
We also introduce $y_3$ such that $1\ll y_3\ll y_2$
and split the integral region into
\begin{eqnarray}
 \int_{y_1}^{y_2}= \int_{y_1}^{y_3} +\int_{y_3}^{y_2} ,~~
 \int_y^{y_2} = \int_y^{y_3} + \int_{y_3}^{y_2} .
\end{eqnarray}
Then Eq.(\ref{i0nu}) is approximated as
\begin{eqnarray}
 I^{1,\nu}(y_1,y_2) \approx {\cal I}_1 + {\cal I}_2 ,
\end{eqnarray}
where
\begin{eqnarray}
 {\cal I}_1 &\!\!=\!\!& 2 \int_{y_1}^{y_3}y dy \int_{y_3}^{y_2}z dz
  \left(
  {(A K_\nu(y)+B I_\nu(y))(C K_\nu(z)+D I_\nu(z))
  \over AD-BC}\right)^2  ,
\\
 {\cal I}_2 &\!\!=\!\!&2 \int_{y_3}^{y_2}y dy \int_{y}^{y_2} z dz
  \left(
  {(A K_\nu(y)+B I_\nu(y))(C K_\nu(z)+D I_\nu(z))
  \over AD-BC}\right)^2 .
\end{eqnarray}
With the relation
\begin{eqnarray}
  AD \ll BC ,~~
  A K_\nu (y) \ll B I_\nu (y) ~~ \textrm{for~large} ~ y
\end{eqnarray}
and modified Bessel function formula given in Appendix, 
the integrals are performed as
\begin{eqnarray}
 {\cal I}_1&\!\!=\!\!&{2\over (AD-BC)^2}
\nonumber
\\
 &&\!\!\!\!\!\!
  \times
  \left[{y^2\over 2}
   ((AK_\nu(y)+BI_\nu(y))^2
    -(AK_{\nu-1}(y)-BI_{\nu-1}(y))(AK_{\nu+1}(y)-BI_{\nu+1}(y)))
 \right]_{y_1}^{y_3}
\nonumber
\\
 &&\!\!\!\!\!\!
  \times
  \left[{z^2\over 2}
   ((CK_\nu(z)+DI_\nu(z))^2
    -(CK_{\nu-1}(z)-DI_{\nu-1}(z))(CK_{\nu+1}(z)-DI_{\nu+1}(z)))
  \right]_{y_3}^{y_2} ,
\nonumber
\\
 &&
\\
 {\cal I}_2
  &\!\!\approx\!\!&2 \int_{y_3}^{y_2}y dy \int_{y}^{y_2} z dz
  \left(
  I_\nu(y)(K_\nu(z)+{D\over C} I_\nu(z))
 \right)^2 .
\end{eqnarray}
For $y_1\to 0$ and $1\ll y_3\ll y_2$, one can show that 
these equations 
become
\begin{eqnarray}
  {\cal I}_1 \to 0 ,
  \qquad
  {\cal I}_2 \approx {1\over 4}\,y_2. 
\end{eqnarray}
Thus we obtain
\begin{eqnarray}
  I^{1,\nu}(0,\Lambda/k) ={1\over 4}~ {\Lambda\over k} .
 \label{wa}
\end{eqnarray}
The value given in Eq.(\ref{wa}) is close to $(1/\pi)(\Lambda/k)$ which is
the number of KK modes below cut off estimated in \cite{Randall:2001gb}.
To this extent,
picture of KK effective theory
works as an intuitive view.
It is also possible to find dependence of the beta function 
on the size of the extra dimension.
If a large cutoff $E=\Lambda$ is introduced 
into analysis in the previous section,
the beta function would have linear dependence on
$n\sim \pi (\Lambda l)$.
Eq.(\ref{wa}) implies that when $kl$ is stabilized
the beta function $b_w$ has linear dependence on $\Lambda l$,
which is similar to KK picture.

\section{Summary and discussions \label{summary}}

We have examined how coupling constants in SU(N) gauge theory
behave at high energy scales, 
depending on the sizes of extra dimensions.
In a flat extra dimension, based on KK picture,
we have solved gauge couplings as functions of the energy
and have present a general equation. 
It has been shown that the gauge couplings 
have logarithmic dependence on the size of the extra dimension
and linear dependence on the number of KK modes.
The running of the gauge coupling can be sensitive to the
number of KK modes once the extra dimension appears with any size.
For example, the first 10 KK modes linearly
would enhance the gauge coupling at an order of magnitude
larger than the scale $\pi/l$.

In order to find some possible patterns 
of the coupling evolution dependent on flavor of bulk and brane 
fermions,
we have also examined the energy when the couplings for 
two gauge group has the same value.
It has been found that 
the energy at a coincident point of the two gauge couplings
is independent of the value of the number of flavor
if the number of flavor for one gauge group is the same as that of 
the other gauge group.
For color $N=2,3$, 
if flavor of bulk fermions satisfies $N_f{}_{N=2}\ge N_f{}_{N=3}$
(with fixed total number of flavor),
there exists such an energy.
And as $N_f{}_{N=3}$ increase, the energy tends to be larger.
This is because the relative sign between bosons and fermions
in the $b_{\textrm{\scriptsize KK}}$ makes the evolution less for 
larger $N_f$.
By our formulation, the energy at a coincident point is
immediately translated into the KK level included 
in the evolution until the coincident point.
As the size of the extra dimension is larger,
the number of KK modes required for coinciding becomes larger.
For $N=2,3$, the number of KK modes required 
is at most of order of ${\cal O}(10)$. 

We have also given an asymptotic solution of
a gauge coupling
in Randall-Sundrum spacetime.
From this solution and 
estimation about
the number of KK modes below cut off in \cite{Randall:2001gb},
picture of KK effective theory
has been found to work to some extent as an intuitive view.
Dependence of couplings on the size of the extra dimension 
has been given with a concept of stabilized $kl$.

It would be interesting to apply our solution and formulation  
in the context of gauge symmetry breaking, grand unification and
supersymmetry.
We have considered the case where all fields have the same KK mass.
It would be straightforward 
to solve gauge couplings in models with field-dependent KK masses.
The value of the gauge couplings itself depends strongly on
field content.
On the other hand,
the number of KK modes required for coinciding of gauge coupling
can be at most ${\cal O}(10)$, depending weakly on field content.
The energy at a coincident point can be 
one to two orders of magnitude larger than $\pi/l$.

\vspace*{10mm}
\begin{center}
{\bf Acknowledgments}
\end{center}  
The author thanks Masud Chaichian for reading the manuscript.
This work is supported by Bilateral exchange programme between 
the Academy of Finland and the Japanese Society for Promotion
of Science.

\begin{appendix}
 \section{Integrals and asymptotic forms of modified Bessel functions}
Integrals of functions including 
the modified Bessel functions $I_\nu$ and $K_\nu$
are shown in Table~\ref{bii}.
\begin{table}[htbp]
\begin{center}
 \caption[]{
Integrals including $I_\nu$ and $K_\nu$.
 \label{bii}}
\begin{tabular}{c|c} \hline\hline
 functions & indefinite integrals \\ \hline
$y I_\nu^2 (z)$ & ${z^2\over 2}(I_\nu^2(z)-I_{\nu-1}(z)I_{\nu+1}(z))$\\
 $y K_\nu^2 (z)$ & ${z^2\over 2}(K_\nu^2(z)-K_{\nu-1}(z)K_{\nu+1}(z))$\\
$y I_\nu(z)K_\nu(z)$ & ${z^2\over 2}(I_\nu(z)K_\nu(z)
    +{1\over 2}(I_{\nu-1}(z)K_{\nu+1}(z)+I_{\nu+1}(z)K_{\nu-1}(z)))$\\
 \hline\hline
\end{tabular}
\end{center}
\end{table}

At the leading order,
$I_\nu$ and $K_\nu$ are independent of $\nu$ 
for large $|z|$,
\begin{eqnarray}
 I_\nu(z)\approx {e^z\over \sqrt{2\pi z}} ,~~
 K_\nu(z)\approx \sqrt{\pi\over 2z}e^{-z} ,
\end{eqnarray}
and for small $z$,
\begin{eqnarray}
 I_\nu (z)\approx \left({z\over 2}\right)^\nu {1\over \Gamma(\nu+1)} ,~~
 K_\nu(z) \approx  
  \left({z\over 2}\right)^{-\nu}
{\Gamma(\nu)\over 2} ,
\end{eqnarray}
for $\nu=n$ ($n$ is non-negative integer)
\begin{eqnarray}
 K_n(z)=K_{-n}(z)
   \approx
  \left\{\begin{array}{l}
   -\log z ~~\qquad\qquad\textrm{for}~ n=0\\
   {1\over 2}(n-1)!\left({z\over 2}\right)^{-n} 
    ~~\textrm{for}~ n\neq 0  .\\ 
	 \end{array}\right. 
\end{eqnarray}

\end{appendix}

\vspace*{10mm}


\end{document}